\documentclass[aps,pra,twocolumn,showpacs]{revtex4}
\usepackage{graphicx,graphics,psfrag,amsmath,calc,amssymb}
\begin{document}

\title{Berezinskii-Kosterlitz-Thouless transition in trapped
quasi-two-dimensional Fermi gas near a Feshbach resonance}
\author{Wei Zhang, G.-D. Lin, and L.-M. Duan}
\affiliation{FOCUS Center and MCTP, Department of Physics, University of Michigan, Ann
Arbor, Michigan 48109}
\date{\today}

\begin{abstract}
We study the superfluid transition in a quasi-two-dimensional Fermi gas with
a magnetic field tuning through a Feshbach resonance. Using an effective
two-dimensional Hamiltonian with renormalized interaction between atoms and
dressed molecules, we investigate the Berezinskii-Kosterlitz-Thouless
transition temperature by studying the phase fluctuation effect. We also
take into account the trapping potential in the radial plane, and discuss
the number and superfluid density distributions. These results can be
compared to experimental outcomes for gases prepared in one-dimensional
optical lattices.
\end{abstract}

\maketitle



\section{Introduction}

\label{sec:introduction}

The study of superfluidity and superconductivity in two-dimensional (2D)
Fermi systems has attracted great attention in the past several decades,
partly because of its close relationship to the problem of high-$T_{c}$
superconductors~\cite{anderson-88, sachdev-03}. Recently, the experimental
progress on creating quasi-low-dimensional atomic gases in optical
lattices~\cite{modugno-03, moritz-05, stoferle-06, chin-06, lewenstein-07}
and on atom chips~\cite{aubin-06, fotagh-07} provides us a possibility of realizing
and studying superfluidity in a more controllable platform. In particular,
with the aid of tuning an external magnetic field through a Feshbach
resonance, the interaction between fermionic atoms can be tuned continuously
from the Bardeen-Cooper-Schrieffer (BCS) limit to the Bose-Einstein condensation
(BEC) limit, so that the BCS-BEC crossover can be studied~\cite{crossover}.

The BCS-BEC crossover in a 2D Fermi system has been discussed at both zero~%
\cite{randeria-90} and finite temperatures~\cite{gusynin-99,
botelho-06}, where an effective 2D Hamiltonian with atom-atom
interaction (called model 1) is employed. However, for strongly
interacting fermionic atoms in a realistic quasi-2D geometry,
due to the inevitable population of many excited levels along
the strongly confined transverse direction~\cite{kestner-06},
one needs to introduce a composite particle called dressed molecule to
account for the population in these transverse excited levels,
and write down a more complicated form for the effective 2D Hamiltonian
(called model 2) with renormalized interaction between atoms
(in the transverse ground level) and dressed molecules~\cite{kestner-07}.
We have shown that the model 1 and the model 2 Hamiltonians lead to
qualitatively distinct results in some cases~\cite{zhang-08}.
For instance, for quasi-2D fermions in a weak global harmonic trap,
the model 1 Hamiltonian predicts a constant value of
the Thomas-Fermi cloud size at zero temperature from the BCS to the BEC
limit. In contrast, the model 2 Hamiltonian predicts that the cloud size
shrinks as one approaches the BEC limit~\cite{zhang-08}, which is the
correct trend as one should expect from the BCS-BEC crossover picture.

In this paper, we extend the discussion of the model 2 Hamiltonian to finite
temperatures by taking into account the fluctuation effects. In particular,
we consider a trapped quasi-2D Fermi gas and study the superfluid transition
temperature across a wide range of Feshbach resonance. Note that according
to the Mermin-Wagner-Hohenberg-Coleman theorem~\cite{MWHC}, phase
fluctuations are dramatic at any finite temperatures such that there will be
no ordinary long-range order in two dimensions. In this case, the superfluid
transition is not accompanied by a Bose condensation, but instead
characterized by a Berezinskii-Kosterlitz-Thouless (BKT) type, for which the
vortex-antivortex pairs start to form at the transition and a topological
order is established~\cite{BKT}. Using the model 2 Hamiltonian, we show
that in a uniform quasi-2D Fermi gas, the BKT transition temperature
$T_{\mathrm{BKT}}$ increases from zero at the BCS limit and approaches
monotonically to almost a constant value about $0.075E_{F}$ at the BEC limit,
where $E_{F}$ is the 2D Fermi energy. Furthermore, we take into account
the weak harmonic trap in the radial plane, and investigate the number and
superfluid density distributions under the local density approximation (LDA).
As a characteristic feature of a BKT transition, a finite jump of superfluid
density is present in the middle of the trap, which takes a universal value
of $2T_{\rm BKT}/\pi $~\cite{nelson-77}.

\section{Formulation}

\label{sec:formulation}

A quasi-2D Fermi gas is typically prepared in experiments by means of a
one-dimensional (1D) deep optical lattice along the axial ($z$) direction,
where tunneling between different lattice sites is completely suppressed. A
1D lattice is described by the potential $V_{\mathrm{ol}}=-V_{0}\exp
[-2(x^{2}+y^{2})/w^{2}]\cos (k_{z}z)$, where $k_{z}$ is the wave vector of
the laser beam, and $w$ is the waist width satisfying $w\gg k_{z}^{-1}$.
This optical lattice potential can be approximated around the minimal points
by a strongly anisotropic pancake-shaped harmonic potential with $V\approx
m(\omega _{z}z^{2}+\omega _{\perp }r^{2})/2$, where $\omega _{z}=\sqrt{%
V_{0}k_{z}^{2}/m}$, $\omega _{\perp }=2\sqrt{V_{0}/(mw^{2})}$, and $r=\sqrt{%
x^{2}+y^{2}}$ is the radial distance. The strong anisotropy of the trap ($%
\omega _{z}\gg \omega _{\perp }$) introduces two well separated energy
scales, which allows us to first deal with the axial degrees of freedom by
deriving an effective 2D Hamiltonian, while the radial degrees of freedom
are left for later treatment with the LDA\ approximation.

The effective 2D Hamiltonian describing a quasi-2D Fermi gas within a strongly
axial confinement is derived in Ref.~\cite{kestner-07}. Under natural units
($\hbar =k_{B}=1$), the Hamiltonian takes the form
\begin{eqnarray}
\mathcal{H} &=&\sum_{\mathbf{k},\sigma }\left( \epsilon _{\mathbf{k}}-\mu
\right) a_{\mathbf{k},\sigma }^{\dagger }a_{\mathbf{k},\sigma }+\sum_{%
\mathbf{q}}\left( \frac{\epsilon _{\mathbf{q}}}{2}+\lambda _{b}-2\mu \right)
d_{\mathbf{q}}^{\dagger }d_{\mathbf{q}}  \notag  \label{eq:tc-Hamiltonian} \\
&+&\frac{\alpha _{b}}{L}\sum_{\mathbf{k},\mathbf{q}}\left( a_{\mathbf{k}%
,\uparrow }^{\dagger }a_{-\mathbf{k}+\mathbf{q},\downarrow }^{\dagger }d_{%
\mathbf{q}}+h.c.\right)   \notag \\
&+&\frac{V_{b}}{L^{2}}\sum_{\mathbf{k},\mathbf{k^{\prime }},\mathbf{q}}a_{%
\mathbf{k},\uparrow }^{\dagger }a_{-\mathbf{k}+\mathbf{q},\downarrow
}^{\dagger }a_{-\mathbf{k^{\prime }},\downarrow }a_{\mathbf{k^{\prime }}+%
\mathbf{q},\uparrow },
\end{eqnarray}
where $a_{\mathbf{k},\sigma }$ and $d_{\mathbf{q}}$ are the annihilation
operators for fermionic atoms and bosonic dressed molecules, respectively.
The dressed molecule consists of a tight pair of atoms distributed in the
transverse (axial) excited levels as well as the population in the closed Feshbach
channel \cite{kestner-07}. In Eq. (1), $\epsilon _{\mathbf{k}}=\mathbf{k}%
^{2}/(2m)$ is the dispersion relation for fermions with mass $m$ and radial
momentum $\mathbf{k}=(k_{x},k_{y})$, $\mu $ is the fermionic chemical potential,
$\sigma =\uparrow ,\downarrow $ labels the pseudo-spin, and $L^{2}$ is the
quantization area. In the remainder of this manuscript, we choose $\omega _{z}$
as the energy unit so that $\mathcal{H}$, $\mu $, and
$\epsilon _{\mathbf{k}}=a_{z}^{2}\mathbf{k}^{2}/2$ become
dimensionless, where $a_{z}\equiv \sqrt{1/(m\omega _{z})}$ is the
characteristic length scale for axial motion.

The 2D effective bare parameters $\lambda _{b}$ (detuning), $\alpha _{b}$
(atom-molecule coupling rate), and $V_{b}$ (background interaction)
in Eq.~\ref{eq:tc-Hamiltonian} are connected with the 2D physical parameters
$\lambda _{p}$, $\alpha _{p}$, and $V_{p}$ through the renormalization
relation
\begin{eqnarray}
\label{eqn:renorm}
\left[ V_{p}^{\text{eff}}(x)\right] ^{-1} &\equiv &\left[ V_{p}+\frac{%
\alpha _{p}^{2}}{x-\lambda _{p}}\right] ^{-1}  \notag \\
&=&\left[ V_{b}+\frac{\alpha _{b}^{2}}{x-\lambda _{b}}\right] ^{-1}+\frac{1}{%
L^{2}}\sum_{\mathbf{k}}\frac{1}{2\epsilon _{\mathbf{k}}+\omega _{z}},
\end{eqnarray}
which is an identity for the variable $x$. The inverse of the physical
effective interaction $\left[ V_{p}^{\text{eff}}(x)\right]^{-1}$ are
determined from the three-dimensional (3D) atomic scattering data
through~\cite{kestner-07}
\begin{equation}
\left[ V_{p}^{\text{eff}}(x)\right] ^{-1}=\frac{\sqrt{2\pi }}{%
a_{z}^{2}}\left[ \left( U_{p}+\frac{g_{p}^{2}}{x-\gamma _{p}}\right)
^{-1}-S_{p}(x)+\sigma _{p}(x)\right] ,  \label{eq:vbneg}
\end{equation}
where $U_{p}=4\pi a_{bg}/a_{z}$, $g_{p}^{2}=\mu _{co}WU_{p}/\omega_{z}$,
and $\gamma _{p}=\mu _{co}(B-B_{0})/\omega _{z}$ are 3D
dimensionless parameters. Here, $a_{bg}$ is the background scattering
length, $\mu _{co}$ is the difference in magnetic moments between the open
and closed collision channels, $W$ is the resonance width, and $B_{0}$ is
the resonance point. The functions in Eq. (\ref{eq:vbneg}) are~\cite%
{kestner-07}
\begin{eqnarray}
S_{p}(x) &\equiv &\frac{1}{\sqrt{32}\pi }\int_{0}^{\infty }ds\left[ \frac{%
\Gamma (s-x/2)}{\Gamma (s+1/2-x/2)}-\frac{1}{\sqrt{s}}\right] ,
\label{eq:sp} \\
\sigma _{p}(x) &\equiv &\frac{\ln \left\vert x\right\vert }{4\pi \sqrt{2\pi }%
},  \label{eq:sigma}
\end{eqnarray}
where $\Gamma (x)$ is the gamma function.

The partition function $\mathcal{Z}$ of the Hamiltonian Eq. (\ref{eq:tc-Hamiltonian})
at temperature $T=\beta ^{-1}$ can be expressed as an
imaginary-time functional integral with action %
\begin{equation}
S=\int_{0}^{\beta }d\tau \left[ \sum_{\mathbf{k,\sigma }}a_{\mathbf{k}%
,\sigma }^{\dagger }\partial _{\tau }a_{\mathbf{k},\sigma }+\sum_{\mathbf{q}%
}d_{\mathbf{q}}^{\dagger }\partial _{\tau }d_{\mathbf{q}}+\mathcal{H}\right]
,  \label{eq:action1}
\end{equation}
where $a_{\mathbf{k},\sigma }=a_{\mathbf{k},\sigma }(\tau )$ and $d_{\mathbf{%
q}}=d_{\mathbf{q}}(\tau )$ become imaginary time dependent variables. By
introducing the Hubbard-Stratonovich field $b_{\mathbf{q}}(\tau )$, which
couples to $a^{\dagger }a^{\dagger }$, and integrating out the fermions, we
obtain $\mathcal{Z}=\int \mathcal{D}[b^{\dagger },b,d^{\dagger },d]\exp (-S_{%
\mathrm{eff}})$, with the effective action given by
\begin{eqnarray}
S_{\mathrm{eff}} &=&\int_{0}^{\beta }d\tau \sum_{\mathbf{q}}\bigg[-\frac{|b_{%
\mathbf{q}}(\tau )|^{2}L^{2}}{V_{b}}-\mu  \notag  \label{eq:action2} \\
&&\hspace{-1.5cm}+d_{\mathbf{q}}^{\dagger }(\tau )\left( \partial _{\tau }+%
\frac{\epsilon _{\mathbf{q}}}{2}+\lambda _{b}-2\mu \right) d_{\mathbf{q}%
}(\tau )\bigg]-\mathbf{Tr}\left[ \ln \mathbf{G}_{\mathbf{k},\mathbf{k}%
^{\prime }}^{-1}\right] .
\end{eqnarray}
Notice that the trace $\mathbf{Tr}$ is taken over momentum, imaginary time,
and Nambu indices of the inverse Nambu matrix $\mathbf{G}^{-1}$, with %
\begin{equation}
\mathbf{G}_{\mathbf{k},\mathbf{k}^{\prime }}^{-1}=\left(
\begin{array}{cc}
-(\partial _{\tau }+\epsilon _{\mathbf{k}}-\mu )\delta _{\mathbf{k},\mathbf{k%
}^{\prime }} & -\Delta _{\mathbf{k}-\mathbf{k}^{\prime }}(\tau ) \\
-\Delta _{\mathbf{k}-\mathbf{k}^{\prime }}^{\dagger }(\tau ) & -(\partial
_{\tau }-\epsilon _{\mathbf{k}}+\mu )\delta _{\mathbf{k},\mathbf{k}^{\prime
}}%
\end{array}%
\right) .  \label{eq:Nambu}
\end{equation}%
Here, $\Delta _{\mathbf{q}}(\tau )\equiv (\alpha _{b}/L)d_{\mathbf{q}}(\tau
)+b_{\mathbf{q}}(\tau )$ is the field whose stationary value corresponds to
the order parameter.

We expand the action around a stationary and homogeneous saddle point
$\Delta _{0}\equiv (\alpha _{b}/L)d_{0}+b_{0}$. The saddle point equation
is derived from the saddle point action $S_{0}[\Delta _{0}]$ via the stationary
conditions $\delta S_{0}/\delta \Delta _{0}=0$. This process leads to
\begin{equation}
\left[ V_{p}^{\text{eff}}(2\mu )\right] ^{-1}=-\frac{1}{L^{2}}\sum_{%
\mathbf{k}}\left[ \frac{1}{2E_{\mathbf{k}}}\tanh \left( \frac{\beta E_{%
\mathbf{k}}}{2}\right) -\frac{1}{2\epsilon _{\mathbf{k}}+1}\right] ,
\label{eq:gapft}
\end{equation}
where $E_{\mathbf{k}}=\sqrt{(\epsilon _{\mathbf{k}}-\mu )^{2}+\Delta _{0}^{2}}$
is the quasi-particle excitation spectrum. We have used the 2D renormalization
relation Eq.~(\ref{eqn:renorm}) in the derivation, which introduces
the term $1/(2\epsilon _{\mathbf{k}}+1)$.

In order to study the fluctuation around the saddle points of $d_{0}$,
$b_{0} $, and $\Delta _{0}$, we next write the field operators as
\begin{subequations}
\label{eqn:orderparameters}
\begin{eqnarray}
d_{\mathbf{q}}(\tau ) &=&\left( |d_{0}|+\delta |d_{\mathbf{q}}(\tau
)|\right) e^{i\phi _{\mathbf{q}}(\tau )}, \\
b_{\mathbf{q}}(\tau ) &=&\left( |b_{0}|+\delta |b_{\mathbf{q}}(\tau
)|\right) e^{i\eta _{\mathbf{q}}(\tau )}, \\
\Delta _{\mathbf{q}}(\tau ) &=&\left( |\Delta _{0}|+\delta |\Delta _{\mathbf{%
q}}(\tau )|\right) e^{i\theta _{\mathbf{q}}(\tau )},
\end{eqnarray}
\end{subequations}
which are all assumed to be slowly varying in both spatial and temporal
coordinates. Here, $\delta |d_{\mathbf{q}}|$, $\delta |b_{\mathbf{q}}|$, and
$\delta |\Delta _{\mathbf{q}}|$ are magnitude fluctuations, while $\phi _{%
\mathbf{q}}$, $\eta _{\mathbf{q}}$, and $\theta _{\mathbf{q}}$ are phase
fluctuations. The magnitude fluctuations correspond to gapped excitations,
whose contribution can be neglected. However, the phase fluctuations
correspond to gapless excitations, and it is their presence that
breaks the ordinary long-range order in two dimensions. Notice that since the
order parameter $\Delta _{\mathbf{q}}$ is a linear combination of $d_{%
\mathbf{q}}$ and $b_{\mathbf{q}}$, neglecting its magnitude fluctuation $%
\delta |\Delta _{\mathbf{q}}|$ is equivalent to assuming a same phase factor
for $d_{\mathbf{q}}$ and $b_{\mathbf{q}}$. We thus have
\begin{equation}
\phi _{\mathbf{q}}(\tau )\equiv \eta _{\mathbf{q}}(\tau )\equiv \theta _{%
\mathbf{q}}(\tau ).  \label{eqn:phase}
\end{equation}

The phase fluctuation $\theta _{\mathbf{q}}(\tau )$ around the saddle points
is in general not a small value, so the phase factor in the order parameter
cannot be directly expanded, but must be treated as a whole. This can be
done by introducing a unitary transformation ${\hat{\mathcal{U}}}\mathcal{O}%
\equiv \exp [i\theta _{\mathbf{q}}(\tau )/2]\mathcal{O}$ for the fields $%
\mathcal{O}=d_{\mathbf{q}}$, $b_{\mathbf{q}}$, and $\Delta _{\mathbf{q}}$.
As a result of this transformation, and after a Fourier transformation back
to the coordinate space, we can expand the effective action around its
stationary point of $\Delta _{0}$, leading to
\begin{equation}
S_{\mathrm{eff}}=S_{0}[\Delta _{0}]+S_{\mathrm{fluc}}[\nabla \theta
,\partial _{\tau }\theta ],  \label{eq:action3}
\end{equation}
where the fluctuation contribution $S_{\mathrm{fluc}}$ contains terms only
involving spatial or temporal gradients of the field $\theta $, which are
assumed to be small.

Expanding $S_{\mathrm{fluc}}$ up to the quadratic order of $\nabla \theta $
and $\partial _{\tau }\theta $ leads to
\begin{equation}
S_{\mathrm{fluc}}\approx \frac{1}{2}\int_{0}^{\beta }d\tau \int d^{2}\mathbf{%
r}\left[ iJ\partial _{\tau }\theta +K(\partial _{\tau }\theta )^{2}+\rho
_{s0}(\nabla \theta )^{2}\right] ,  \label{eq:action4}
\end{equation}
where $\rho _{s0}$ represents the phase stiffness (or called the superfluid
density) and takes the form
\begin{eqnarray}
\rho _{s0} &=&\frac{1}{4m}\Bigg\{2 n_{b}+\frac{1}{L^{2}}\sum_{\mathbf{k}}\left[
1-\frac{\epsilon _{\mathbf{k}}}{E_{\mathbf{k}}}\tanh \left( \frac{\beta E_{%
\mathbf{k}}}{2}\right) \right]  \notag  \label{eq:rhos} \\
&&\hspace{1cm}-\frac{\beta \hbar ^{2}}{4mL^{2}}\sum_{\mathbf{k}}k^{2}\mathrm{%
sech}^{2}\left( \frac{\beta E_{\mathbf{k}}}{2}\right) \Bigg\}.
\end{eqnarray}
Here, $n_{b}=\Delta ^{2}\partial \lbrack V_{p}^{\text{eff}}(2\mu
)]^{-1}/\partial \mu $ denotes the dressed-molecular population. Other
parameters in Eq. (\ref{eq:action4}) are
\begin{eqnarray}
J &=& 2 n_{b}-\frac{1}{L^{2}}\sum_{\mathbf{k}}\frac{\epsilon _{\mathbf{k}}-\mu
}{E_{\mathbf{k}}}\tanh \left( \frac{\beta E_{\mathbf{k}}}{2}\right) ,
\label{eq:K} \\
K &=&\frac{m}{8\pi }\left[ 1+\frac{\mu }{\sqrt{\mu ^{2}+\Delta _{0}^{2}}}%
\tanh \left( \frac{\beta \sqrt{\mu ^{2}+\Delta _{0}^{2}}}{2}\right) \right] .
\end{eqnarray}
\begin{figure}[tbph]
\includegraphics[width=8.5cm]{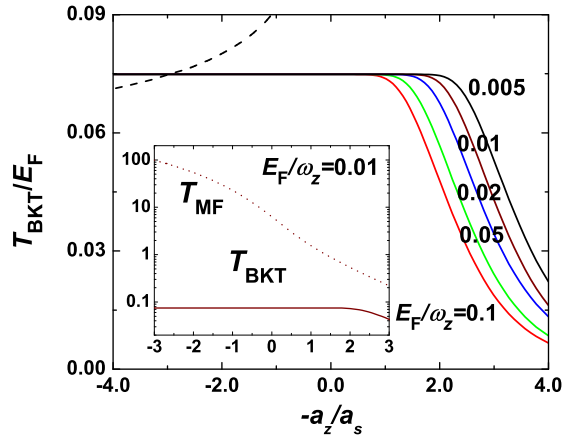}
\caption{(Color online) Superfluid transition temperature $T_{\mathrm{BKT}}$
(solid lines) of a uniform quasi-2D Fermi gas for various values of $E_{F}/%
\protect\omega _{z}$. Notice that $T_{\mathrm{BKT}}$ increases monotonically
from the BCS (right) to the BEC (left) side of the resonance, and approaches
to a limiting value of $\sim 0.075E_{F}$. In the BEC limit, $T_{\rm BKT}$ can also
be estimated by considering a weakly interacting gas of composite bosons with
3D scattering length of $0.6 a_s$, leading to a logarithmically
varying result (dashed line). The mean-field transition temperature
$T_{\mathrm{MF}}$ (dashed line in the inset) is also shown as a comparison
for $E_{F}/\protect\omega_{z}=0.01$.}
\label{fig:homogeneous}
\end{figure}

By decomposing the phase fluctuation $\theta (\mathbf{r},\tau )=\theta _{v}(%
\mathbf{r},\tau )+\theta _{\mathrm{sw}}(\mathbf{r},\tau )$ into a vortex
part $\theta _{v}$ and a vortex-free spin-wave part $\theta _{\mathrm{sw}}$,
one can separate the effective action $S_{\mathrm{eff}}$, and hence the
thermodynamic potential as~\cite{botelho-06}
\begin{equation}
\Omega =\frac{S_{\mathrm{eff}}}{\beta }=\Omega _{0}+\Omega _{v}+\Omega _{%
\mathrm{sw}},  \label{eqn:action5}
\end{equation}
where $\Omega _{0}$, $\Omega _{v}$ and $\Omega _{\mathrm{sw}}$ are the
corresponding contributions from the saddle point action $S_{0}$, the vortex
contribution from $\theta _{v}$, and the spin-wave contribution from $\theta
_{\mathrm{sw}}$, respectively. The spin-wave part can be integrated out to
give $\Omega _{\mathrm{sw}}=T\sum_{\mathbf{k}}\ln \left( 1-e^{-ck/T}\right) $%
, where $c=\sqrt{\rho _{s0}/J}$ is the speed of the spin-wave. The number
equation hence can be obtained via $N=-\partial \Omega /\partial \mu $,
leading to
\begin{eqnarray}
n &=&\frac{N}{L^{2}}=n_{\mathrm{sw}} + 2 n_{b}  \notag  \label{eq:number1} \\
&&+\frac{1}{L^{2}}\sum_{\mathbf{k}}\left[ 1-\frac{\epsilon _{\mathbf{k}}}{E_{%
\mathbf{k}}}\tanh \left( \frac{\beta E_{\mathbf{k}}}{2}\right) \right]
\end{eqnarray}
with the spin-wave contribution $n_{\mathrm{sw}}=-\partial \Omega _{\mathrm{%
sw}}/\partial \mu $. By writing the number equation as above, we implicitly
assume a vanishingly small vortex density, such that the vortex contribution
to the number equation can be safely neglected. This condition is in general
valid for temperatures around or below the superfluid transition
temperature~\cite{minnhagen-87}.
\begin{figure}[tbph]
\includegraphics[width=8.5cm]{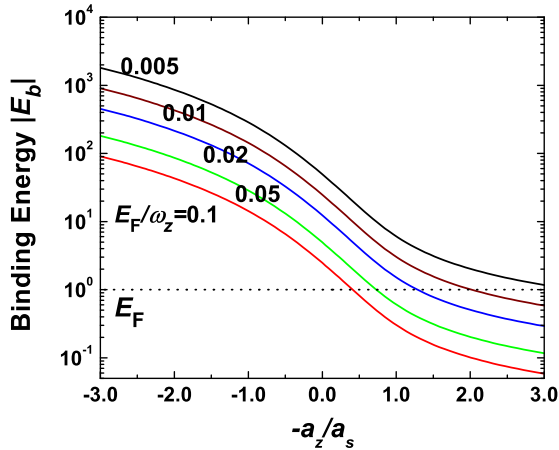}
\caption{(Color online) Two-body binding energy $|E_{b}|$ (solid lines) as a
function of $a_{z}/a_{s}$ for various values of $E_{F}/\protect\omega _{z}$.
Notice that $|E_{b}|$ can exceed $E_{F}$ (dotted) even on the high-field
side of the resonance point, hence pushes the system into the BEC limit as
indicated by the transition temperature shown in Fig.~\ref{fig:homogeneous}.}
\label{fig:bindE}
\end{figure}

The $\rho _{s0}$ is the superfluid density at the microscopic scale, and the
vortex fluctuation tends to renormalize (decrease) the superfluid density
when the physical scale $l$ goes up. When the superfluid density is
renomalized to zero with $l\rightarrow \infty $, the
Berezinskii-Kosterlitz-Thouless (BKT) transition takes phase where the
system changes from a superfluid phase to a normal phase. The
renormalization-group (RG) flow depends on the vortex core energy $E_{c}$,
which can be estimated by $E_{c}=0.78 \pi \rho _{s0}$~\cite{minnhagen-85}.
The RG equations for the superfluid density $\rho _{s}$ and the fugacity
$y=e^{-\beta E_{c}}$ are given by~\cite{kosterlitz-74}
\begin{eqnarray}
\frac{dK^{-1}(l)}{dl} &=&4\pi ^{3}y^{2}(l)+O(y^{3}),  \label{eqn:RG1} \\
\frac{dy(l)}{dl} &=&\left[ 2-\pi K(l)\right] y(l)+O(y^{2}),  \label{eqn:RG2}
\end{eqnarray}
where $K(l)\equiv \rho _{s}(l)/T$. The initial conditions are $\rho
_{s}(l=0)=\rho _{s0}$ and $y(l=0)=\exp (-\pi \rho _{s0}/2T)$. The fixed
point $(y(\infty ),K(\infty ))=(0,2/\pi )$ of these RG equations corresponds
to the critical condition for the BKT transition, where the vortex-
antivortex pairs start to dissociate and break superfluidity. Thus, the
equation for the critical temperature $T_{c}=T_{\mathrm{BKT}}$ is determined
as~\cite{kosterlitz-74}
\begin{equation}
T_{\mathrm{BKT}}=\frac{\pi }{2}\rho _{s}^{R},  \label{eq:TBKT}
\end{equation}
where $\rho _{s}^{R}=\lim_{l\rightarrow \infty }\rho _{s}(l)$ is the
renormalized superfluid density. This equation must be solved together with
the saddle-point equation (\ref{eq:gapft}), the number equation (\ref%
{eq:number1}), and the RG equations (\ref{eqn:RG1}) and (\ref{eqn:RG2}) to
determine $\mu $, $\Delta _{0}$, and $T_{\mathrm{BKT}}$ as functions of the
magnetic field detuning. As a comparison, the mean-field transition
temperature $T_{\mathrm{MF}}$ can be calculated by solving only Eqs. (\ref%
{eq:gapft}) and (\ref{eq:number1}) with $\Delta _{0}=0$. Next, we will
consider some specific systems and discuss the corresponding numerical
results.

%

\section{Numerical results}

\label{sec:results}

We first ignore the trapping potential in the radial plane, and consider a
uniform quasi-2D Fermi gas. A typical set of results of $T_{\mathrm{BKT}}$
for various values of $\omega_z$ are shown in Fig.~\ref{fig:homogeneous}.
Here, we consider the case of $^6$Li and use the parameters $a_{bg} = -1405
a_0$, $W = 300$ G, and $\mu_{co} = 2 \mu_B$, where $a_0$ and $\mu_B$ are
Bohr radius and Bohr magneton, respectively. We also notice that when
plotted as functions of the inverse 3D scattering length $a_z/a_s$, the
results for $^{40}$K are very close to those for $^6$Li, indicating the
universal behavior around the resonance point.

Notice that the superfluid transition temperature $T_{\mathrm{BKT}}$
increases continuously from the BCS to the BEC side of the Feshbach
resonance, and saturates to a limiting value of $T_{\mathrm{BKT}}\approx
0.075E_{F}$. This limiting value of $T_{\mathrm{BKT}}$ is expected for $\rho
_{s0} \approx n/(4m)$, which corresponds to the BEC limit of a quasi-2D
Fermi gas where paired fermions behave like weakly interacting bosons with
number density $n/2$ and mass $2m$. In the BEC limit, we can also consider the
particles as composite bosons with effective three-dimensional scattering length
of $0.6a_{s}$~\cite{petrov-04}. For weakly interacting bosons, more accurate
results are available for the BKT transition temperature based on a combination
of the quantum Monte Carlo simulation~\cite{prokofev-01} and the renormalization
group approach~\cite{holzmann-07}, which predicts that $T_{\mathrm{BKT}}$ goes
down logarithmically with the scattering length. This more accurate result
in the BEC limit is also shown in Fig.~\ref{fig:homogeneous}, which is slightly
deviant from our fermionic calculation.

We note also from Fig.~\ref{fig:homogeneous} that the BEC limit value of $T_{%
\mathrm{BKT}}$ can even be reached on the BCS side of the Feshbach resonance
(with $a_{s}<0$). In fact, since a two-body bound state is always present in
quasi-low dimensions at arbitrary detunings, the BEC limit of a quasi-2D
Fermi gas can be realized as long as the condition $|E_{b}|\gg E_{F}$ is
satisfied, regardless of the sign of the 3D scattering length $a_{s}$. This
property is in clear contrast to the 3D case, where a bound state is only
present with positive $a_{s}$, hence the BEC limit can only be realized on
the low-field side of a Feshbach resonance. Here, the two-body binding
energy $|E_{b}|$ is determined by solving the Hamiltonian~(\ref{eq:tc-Hamiltonian})
for two particles, which gives
\begin{equation}
\left[ V_{p}^{\text{eff}}(E_{b})\right] ^{-1}=\frac{1}{4\pi
a_{z}^{2}}\ln (-E_{b}).  \label{eq:Eb}
\end{equation}
The corresponding results for $|E_{b}|$ are plotted in Fig.~\ref{fig:bindE}.
Notice that for the case with $E_{F}/\omega _{z}=0.005$, the binding energy $%
\vert E_{b} \vert \approx 2.0E_{F}$ for $a_{z}/a_{s}\approx -2.0$, leading to a BKT
transition temperature of the BEC limit value $T_{\mathrm{BKT}}\approx
0.075E_{F}$, as shown in Fig.~\ref{fig:homogeneous}.

Another feature of Fig.~\ref{fig:homogeneous} is that the transition
temperature $T_{\mathrm{BKT}}$ increases with the axial trapping frequency $%
\omega _{z}$ at a fixed detuning. This trend can also be easily understood
by analyzing the two-body binding energy $|E_{b}|$. In fact, at a given
detuning, $|E_{b}|$ increases with $\omega _{z}$ as shown in Fig.~\ref%
{fig:bindE}, hence pushes the quasi-2D system further to the BEC limit.
\begin{figure}[tbp]
\includegraphics[width=8.5cm]{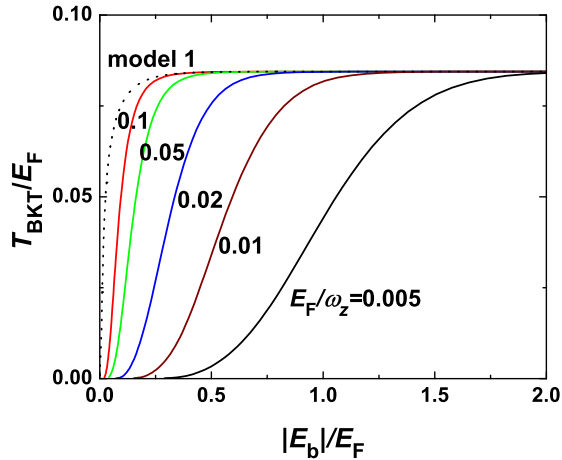}
\caption{(Color online) Superfluid transition temperature $T_{\mathrm{BKT}}$
(solid lines) plotted as functions of the two-body binding energy $|E_{b}|$.
The results are compared with the outcome of an effective 2D Hamiltonian
with renormalized atom-atom interaction (model 1, dotted line), which is
discussed in Refs.~\protect\cite{gusynin-99,botelho-06}. }
\label{fig:homogeneous-Eb}
\end{figure}

The results of $T_{\mathrm{BKT}}$ should also be compared with the outcome
of an effective 2D Hamiltonian with renormalized atom-atom interaction
(model 1), which is discussed in Refs.~\cite{gusynin-99,botelho-06}. In Fig.~%
\ref{fig:homogeneous-Eb}, we show $T_{\mathrm{BKT}}$ calculated using both
models as functions of the binding energy $|E_{b}|$. One prediction of the
model 1 is that the many-body physics (such as the BKT transition
temperature) takes a universal behavior, which only depends on the two-body
binding energy $|E_{b}|$ in the unit of $E_{F}$ \cite{randeria-90,gusynin-99,botelho-06}.
As shown in Fig. \ref{fig:homogeneous-Eb}, this is clearly not the case for the results
from the model 2, where the transition temperature $T_{\mathrm{BKT}}$ also depends on
the other energy scale such as the transverse trapping frequency $\omega _{z}$.
Notice that both of the models predict roughly the same limiting value of
$T_{\mathrm{BKT}}$ for large values of $|E_{b}|$ (in the BEC limit),
since in that limit the system behaves like weakly
interacting bosons, and $T_{\mathrm{BKT}}$ gets very insensitive to the
interaction between these composite bosons (logarithmic dependence as
mentioned above).
\begin{figure}[tp]
\includegraphics[width=9.0cm]{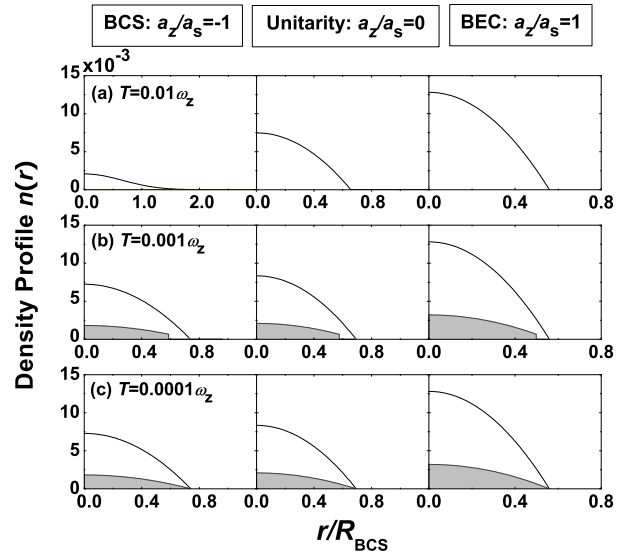}
\caption{(Color online) In-trap number (solid) and superfluid (shaded area)
density distributions at (a) $T=0.01\protect\omega _{z}$, (b) $T=0.001\protect%
\omega _{z}$, and (c) $T=0.0001\protect\omega _{z}$. For each temperature,
three typical detunings are considered with $a_{z}/a_{s}=-1$ (BCS, left), $%
a_{z}/a_{s}=0$ (unitarity, middle), and $a_{z}/a_{s}=1$ (BEC, right),
respectively. Notice the finite jump of superfluid density in the middle of
the trap at intermediate temperatures. The plots are made for $^{6}$Li,
while the results for $^{40}$K are similar. Parameters used in this plot are
$\protect\omega _{z}=2\protect\pi \times 62$ kHz, $\protect\omega _{\perp }=2%
\protect\pi \times 10$ Hz, $N=10000$, and $R_{\mathrm{BCS}}=\protect\sqrt{2%
\protect\omega _{z}/\protect\omega _{\perp }}(N)^{1/4}a_{z}$ is the
zero-temperature Thomas-Fermi cloud size of a quasi-2D ideal Fermi gas with
particle number $N$. }
\label{fig:finiteTintrap}
\end{figure}

After discussing the case of a uniform quasi-2D Fermi gas, we next consider the
trapping potential in the radial plane $V_{\perp }(r)=m\omega _{\perp
}r^{2}/2$. Under the local density approximation (LDA), we consider a
position dependent chemical potential $\mu (r)=\mu _{0}-V_{\perp }(r)$.
Here, $\mu _{0}$ is the chemical potential at the trap center, which must be
determined by fixing the total particle number in the trap $N=2\pi
\int_{0}^{\infty }n(r)rdr$. Using this technique, we calculate the in-trap
number and superfluid density distributions in the radial plane, which are
shown in Fig.~\ref{fig:finiteTintrap}. From the left to the right, results
are sequentially presented for the BCS side of the resonance ($a_{z}/a_{s}=-1
$), unitarity ($a_{z}/a_{s}=0$), and the BEC side of the resonance ($%
a_{z}/a_{s}=1$). For each case, temperature is varied within two orders of
magnitude, showing that superfluidity is absent at high temperatures and
starts to build from the trap center as decreasing $T$. At low enough
temperatures, superfluidity nearly extends to the whole trap, and the number
density profile approaches the zero temperature results~\cite{zhang-08}.
Furthermore, notice that there is a finite jump of
superfluid density present in the middle of the trap at intermediate
temperatures, as shown in panels of Fig.~\ref{fig:finiteTintrap}b. This
discontinuity in $\rho_{s}^{R}$ signatures the position where the
superfluid transition takes place, and takes a universal value of $\delta
\rho _{s}^{R}=2T/\pi $ characterizing a phase transition in the BKT
universality class~\cite{nelson-77}.


\section{Conclusion}

\label{sec:conclusion}

In summary, we have studied the BKT type of superfluid transition of a
quasi-2D Fermi gas based on an effective 2D Hamiltonian with renormalized
interaction between atoms and dressed molecules. The finite temperature
effect has been taken into account by incorporating phase fluctuations over
the saddle point solution. Using the effective Hamiltonian that we derived
before with the explicit parameters \cite{kestner-07}, we establish the BKT\
transition temperature as a function of the 3D atomic scattering length,
making it possible to directly compare the result with the experimental
measurements. We also compare the predictions from this effective
Hamiltonian (model 2) with the results from model 1 (where the 2D
Hamiltonian is described by atom-atom interaction with an effective
scattering length~\cite{gusynin-99,botelho-06}), and they differ
significantly in certain parameter regions. In particular, the universal
behavior predicted by the mode l (many-body physics depends only on the two-body
binding energy) is not the case for the model 2, where the BKT transition
temperature also has explicit dependence on the transverse trapping
frequency $\omega _{z}$. This difference can be quantitatively tested by future
experiments. Under the local density approximation, we have also calculated the
in-trap number and superfluid density distributions at finite temperatures,
which can be compared with experimental results in a harmonic trap.

\acknowledgements

This work was supported under the MURI program and under ARO Grant No. W911NF0710576
with funds from the DARPA-OLE Program.


\end{document}